\documentclass[prb,preprint]{revtex4-1}
\usepackage{float}
\usepackage[caption = false]{subfig}
\usepackage[final]{graphicx}

\usepackage{dcolumn}
\usepackage{bm}
\usepackage{amssymb,amsmath}
\usepackage{color}
\usepackage{epstopdf}

\begin{document}
	
	\title{Emergent structural length scales in a model binary glass --- the micro-second molecular dynamics time-scale regime}
	\author{P. M. Derlet}
	\email{Peter.Derlet@psi.ch} 
	\affiliation{Condensed Matter Theory Group, Paul Scherrer Institut, CH-5232 Villigen PSI, Switzerland}
	\author{R. Maa{\ss}}
	\affiliation{Department of Materials Science and Engineering and Frederick Seitz Materials Research Laboratory, University of Illinois at Urbana-Champaign, 1304 West Green Street, Urbana, Illinois 61801, USA}
	\date{\today}
	
\begin{abstract}
It is now possible to routinely perform atomistic simulations at the microsecond timescale. In the present work, we exploit this for a model binary Lennard-Jones glass to study structural relaxation at a timescale spanning up to 80 microseconds. It is found that at these longer time-scales, significant mobility and relaxation occurs, demonstrating a reproducible relaxation trajectory towards a class of asymptotic structural states which are strongly heterogeneous. These structures are discussed both in terms of the ideal glass state and a production route towards novel amorphous crystalline nano-composite micro-structures with strong interface stability. 
\end{abstract}
\maketitle
\section{Introduction} \label{sec-intro}

Bulk metallic glasses are strongly out-of-equilibrium industrially relevant materials with exceptional physical and mechanical properties~\cite{Hufnagel2016}. Produced by a rapid quench from the under-cooled liquid, the amorphous state emerges during the glass transition temperature ($T_{\mathrm{g}}$) regime with an extensive configurational entropy. However, this entropy is considerably lower than that of the liquid indicating strong constraints on the available atomic configurations~\cite{Debenedetti2001}. Understanding these constraints and how they are overcome is central to a proper control of both structure and mechanical properties --- a parameter space that is explored via structural relaxation and rejuvenation of the glass. Increasingly, such structural tuning is assessed via the excess enthalpy storage of the glass, which is amongst the highest of any metallic material~\cite{Sun2016,Kuechemann2017,Kuechemann2018,Ding2019}.

Two structural asymptotes have remained omnipresent in both the experimental and theoretical efforts to understand glass tunability. The first is crystallization due to annealing protocols close to $T_{\mathrm{g}}$, and the second is relaxation towards the ideal glass state, characterized by a sub-extensive configurational entropy~\cite{Kauzmann1948}. Generally these are considered to be quite different and unrelated structural states. Due to undesirable mechanical properties and experimental difficulties, as-produced glasses tend to be away from the latter extreme, whereas metallic-glass matrix composites (multi-phase structures containing crystalline and possibly amorphous regions) embrace the former. On the other hand, approaching the ideal glass limit, the low temperature limit of the undercooled liquid, poses interesting theoretical questions as first pointed out by Kauzmann~\cite{Kauzmann1948} and also the alluring possibility of a glass structure which is in a true meta-equilibrium state.

Once a glassy structure is obtained after quenching a melt, there will be a natural tendency for the amorphous solid to relax. The resulting structural evolution is seen in small changes in the pair distribution function (PDF) at radial distances beyond the nearest neighbour peak(s) that characterize the short-range order (SRO). The more distant order has been referred to as medium-range order (MRO). A common framework to understand the developing MRO is with respect to the packing of SRO icosahedral content~\cite{Sheng2006}. An atom is said to be icosahedrally coordinated if its nearest neighbour spatial environment can be characterized by a Wigner-Seitz or Voronoi generated twelve faced polyhedron where each face consists of three edges --- the icosahedron. This originates from the more fundamental unit structure of the tetrahedron~\cite{Chaudhari1978}. When isolated, this atomic arrangement can usually minimize the energy of all six bonds. However it is not possible to pack such tetrahedra in a space-filling way whilst maintaining this minimum ground-state energy and therefore bonds become frustrated, both in terms of energy and volume. Optimal space-filling packing of icosahedra at the length-scale of the MRO (and beyond) is a difficult problem involving defected icosahedra and more general Voronoi polytopes~\cite{Frank1952,Nelson1983a,Nelson1983b}, underlying the very close connection between local structure, bond energy and atomic volume~\cite{Ding2017}.

The relationship between dynamical heterogeneities and structure in the under-cooled liquid-metal regime gives additional insight into the nature of this structural relaxation. In particular the connection to amorphous disorder, which is normally considered to have little in common with those local structures ultimately giving rise to crystalline order~\cite{Charbonneau2016}. This aspect may be articulated as the competition of local structures characterized by five- (icosahedral) and six-fold bond symmetries (close-packed)~\cite{Taffs2016}. Common equilibrium crystal structures are close-packed and therefore dominated by six-fold bond-order. However there do exist a class of crystal structures for which the five-fold bond symmetry dominates. These are the so-called topologically close packed phases discovered by Frank and Kasper~\cite{Frank1958}. For binary systems common realizations are the cubic (C15), the two hexagonal (C14 and C36) laves structures, and the A15 structure. Atomistic simulations of model binary glasses have revealed connected fragments of the Laves C15 structure for both Lennard-Jones (LJ) force models~\cite{Pedersen2010} (using the Wahnstr{\o}m parameterization~\cite{Wahnstrom1991}), and material specific force models for the experimentally realizable CuZr binary alloy~\cite{Zemp2014,Zemp2016,Ryltsev2016}. Experimental evidence for such local icosahedral structures mainly relies on a combination of x-ray diffraction and the reverse Monte-Carlo method. For the CuZr alloy, for a range of concentration, this method suggests extended structures of connected icosahedral units~\cite{Wang2008} possibly inter-dispersed amongst liquid-like structures~\cite{Li2009}. For the case of the Cu$_{50}$Zr$_{45}$Al$_{5}$ ternary alloy, reverse Monte Carlo simulations in combination with fluctuation electron microscopy suggested the existence of a population of both icosahedral and crystal-like (six-fold bonded) atomic structures~\cite{Hwang2012}. The LJ work of Pedersen {\em et al}~\cite{Pedersen2010} found that such Laves fragments took the form of two dimensional ring-structures referred to as Frank-Kasper backbone polyhedrons. Earlier work also showed that at sufficiently high temperatures, within the undercooled liquid regime, the Wahstr{\o}m LJ potential can also crystallize into the C15 Laves phase~\cite{Pedersen2006}.

Past MD works probing the nano-second timescale~\cite{Ding2014,Derlet2017,Derlet2017a,Derlet2018}, have shown that below $T_{\mathrm{g}}$, the cohesive energy decreases with respect to increasing icosahedral content. Indeed, Ref.~\cite{Derlet2018}, which considers the model Wahnstr{\o}m LJ binary glass, has demonstrated that glass relaxation in terms of both energy and pressure/volume originates directly from the conversion of non-icosahedral environments to icosahedral environments, via structural excitations characterized by thermally activated collective string-like atomic displacements~\cite{Derlet2017,Derlet2017a}. Such a geometry of structural evolution has also been observed in the dynamical heterogeneities of the under-cooled liquid~\cite{Donati1998,Schroeder2000,Gebremichael2004,Vogel2004,Kawasaki2013}.

In the present work we employ atomistic simulation to investigate the non-negligible structural relaxation in a model binary glass which occurs at time-scales up to approximately $80$ $\mu$sec. We show that that the linear relation between icosahedral content and energy remains valid over this time-scale and therefore to glassy structures significantly more relaxed than previously considered. We find that this very general relaxation trajectory can lead to emergent structural length-scales reflecting both the nano-composite and ideal-glass structural asymptotes, suggesting that the ideal glass and crystallization might be intimately related if not associated with the same structural state. We argue that this perspective offers a new approach to structural tuning of nano composite (nc) amorphous-crystalline microstructures, which we here reveal for binary metallic glass formers. In sec.~\ref{sec-simulation} we outline our simulation approach, in sec.~\ref{sec-results} we detail the major simulation results, and in sec.~\ref{sec-discussion} we discuss them within the framework of temperature-time-transition (TTT) diagrams that are now accessible within the time-frame of an atomistic simulation. Finally, we formulate a number of open questions resulting from this work and in sec.~\ref{sec-conc} we summarize and conclude.

\section{Simulation Methodology} \label{sec-simulation}

As already noted, the physics of structural frustration can be captured by simple pair potential force models such as the Lennard-Jones (LJ) potential~\cite{Rodney2011}, whereas for metal-specific applications embedded atom~\cite{Daw1984} or second-moment~\cite{Finnis1985} model potentials may be used. This demonstrates the more involved force models are not essential to understanding the underlying structural glass physics~\cite{Derlet2017a,Derlet2018} --- glassy physics is a universal phenomenon~\cite{Debenedetti2001,Kauzmann1948,Goldstein1969} which can be equally well simulated using an appropriate LJ parameterization. Because of this, and due to their more-rapid numerical evaluation, we use a model binary LJ system parameterized by Wahnstr\"{o}m~\cite{Wahnstrom1991} to investigate its microstructural relaxation behaviour at the timescale of many tens of $\mu$sec. For two atoms of type $a$ and $b$, separated by a distance $r$, the interaction potential energy is given by,
\begin{equation}
V_{ab}(r)=4\varepsilon\left(\left(\frac{\sigma_{ab}}{r}\right)^{12}-\left(\frac{\sigma_{ab}}{r}\right)^{6}\right),
\end{equation}
where $\sigma_{22}=5/6\sigma_{11}$ and $\sigma_{12}=\sigma_{21}=11/12\sigma_{11}$. The atoms of type 1 may be considered as the larger atom type. The atomic masses of the two atom types are arbitrarily chosen such that $m_{1}/m_{2}=2$. For a molecular dynamics iteration, a time step of $0.002778\sigma_{11}\sqrt{m_{1}/\varepsilon}$ is used. The distance unit is taken as $\sigma=\sigma_{11}$ and the energy unit as $\varepsilon$. For this work, the potential is truncated to a distance 2.5$\sigma$. Atoms of type 1 may be considered as the larger atom. All molecular dynamics (MD) simulations are done using the LAMMPS software~\cite{LAMMPS}, and atomic scale analysis via the OVITO visualization software~\cite{Stukowski2010}.

When metallic units (representative of say, CuZr) are taken for $\varepsilon$ and $\sigma$, an MD time-step is of the order of a femto-second. Thus one billion MD iterations corresponds to approximately one micro-second. Throughout the remainder of this paper, simulation times will be measured with respect to one billion MD iterations, which in turn is approximated as one micro-second.

A model binary glass system is created via an NVT temperature quench from a close-to-equilibrium liquid state~\cite{Derlet2017,Derlet2017a}. Following either pressure or cohesive energy, a change in slope with respect to temperature indicates a glass transition is reached and the amorphous solid regime is entered. The temperature at which this occurs is referred to as the fictive glass transition temperature ($T_{\mathrm{f}}$). For MD, $T_{\mathrm{f}}$ is usually taken as a reference glass transition temperature, $T_{\mathrm{g}}$. Experimentally, $T_{\mathrm{f}}$ is seen as the glass transition temperature without any additional annealing protocol. Such a temperature protocol, sometimes with a pressure protocol, embodies the standard method to produce a glass using atomistic simulation~\cite{Rodney2011}.  At a somewhat lower temperature than $T_{\mathrm{f}}$, the temperature quench may be halted and long-time constant temperature simulations are performed, followed by a further quench to zero temperature. This annealing step is known to produce more relaxed samples~\cite{Derlet2017a}. The present work considers only fixed volume simulations, the so-called NVT ensemble. Recent work has shown that when performed at appropriately rescaled temperatures, both fixed volume and fixed pressure (NPT ensemble) sample production protocols result in similar relaxation trajectories and thus similar final structural states with zero temperature volumes that differ by only $\sim0.2\%$. See Ref.~\cite{Derlet2018} for more details.

In the first instance, we consider a glass sample consisting of 500 atoms, at a fixed volume per atom equal to 0.767$\sigma^{3}$. This sample is initially produced using the linear-quench protocol just discussed. Such a small atom number is chosen to maximize computational throughput, but is large enough to accommodate the local structural excitations (LSEs) that mediate general structural relaxation~\cite{Swayamjyoti2014,Derlet2017,Derlet2017a}. The precise details of the quench protocol are identical to Ref.~\cite{Derlet2017a}, resulting in the present sample having the same value of $T_{\mathrm{f}}$ as that of Ref.~\cite{Derlet2017a}.

\section{Results} \label{sec-results}

\subsection{Long time $T<T_{\mathrm{f}}$ relaxation} \label{ssec-lts}

\begin{figure*}
	\subfloat[]{\includegraphics[width=0.4\linewidth]{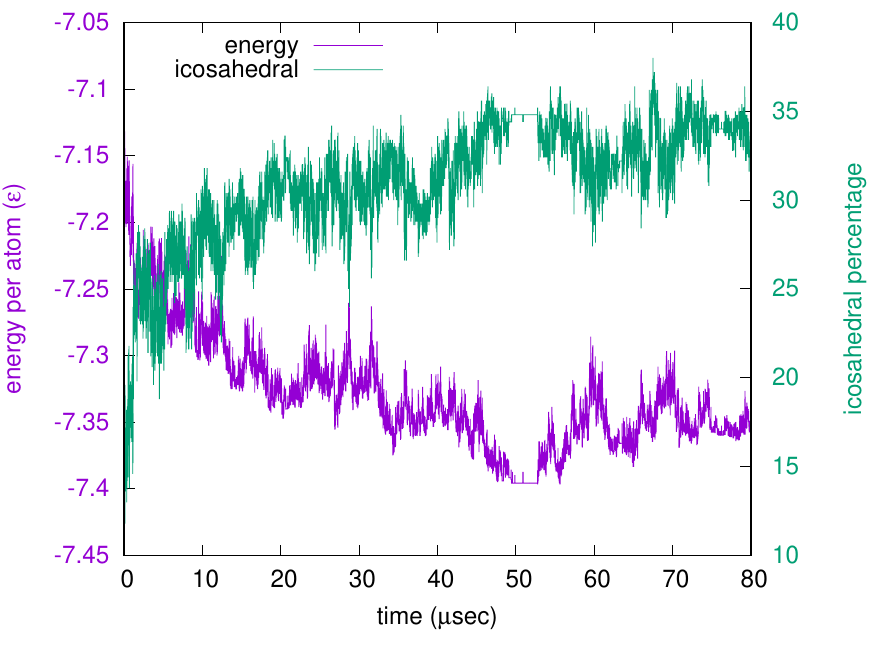}}
	\subfloat[]{\includegraphics[width=0.4\linewidth]{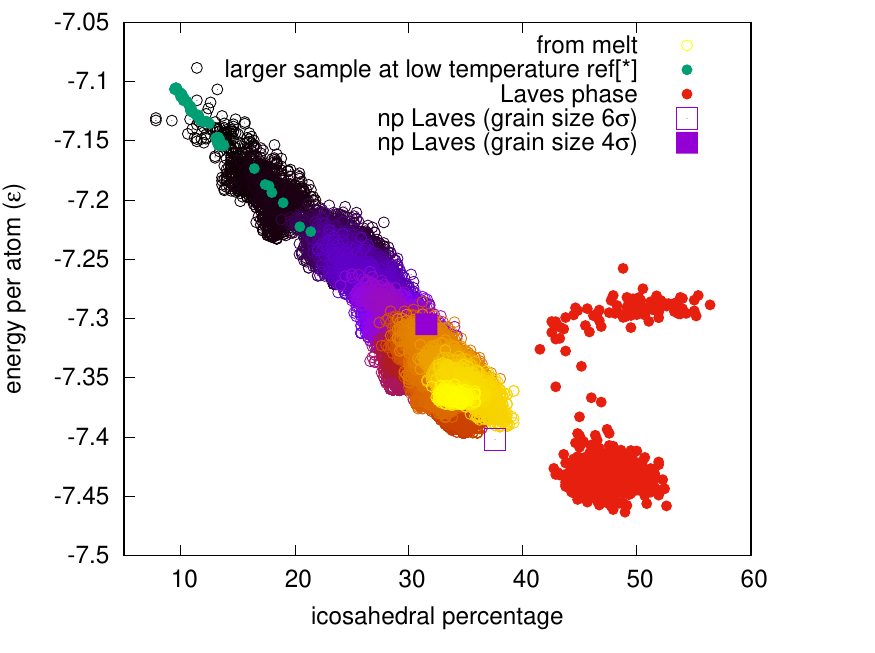}} \
	\subfloat[]{\includegraphics[width=0.8\linewidth,trim=1.5cm 1cm 1cm 1cm,clip]{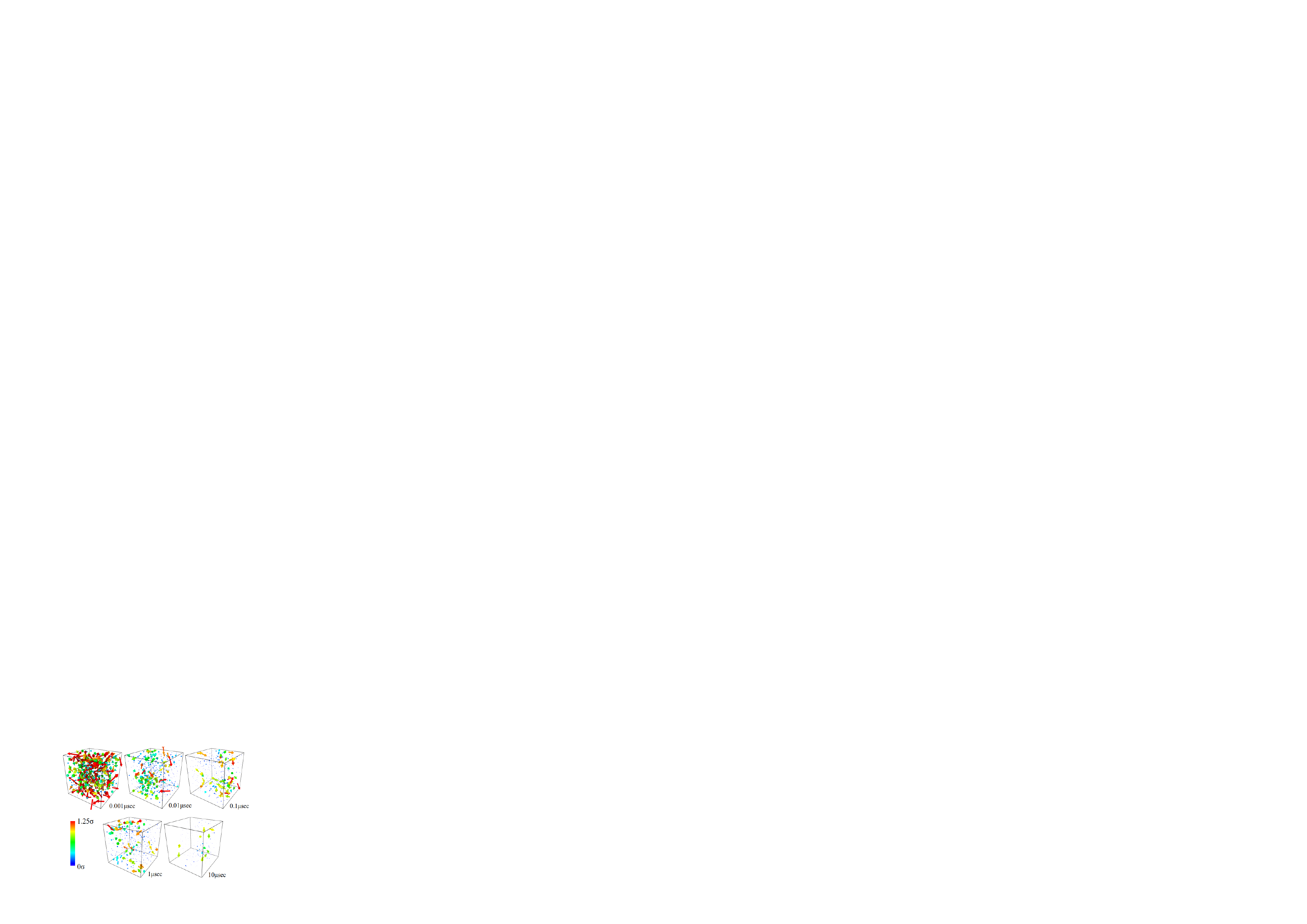}}
	\caption{a) Energy per atom and icosahedral content evolution over a time period of 80 $\mu$sec at an annealing temperature $0.95T_{\mathrm{f}}$. b) Scatter plot of energy per atom versus icosahedral content. Also shown is data for a larger sample of 32000 atoms~\cite{Derlet2017a}, a geometrically constructed nano-composite Laves phase with two characteristic grain sizes, and the results of a hybrid-MD/MC simulation starting from a chemically disordered Laves crystal. c) Atomic displacement maps between configuration separated by 1 million MD iterations (0.001 $\mu$sec) taken at 0.001, 0.01, 0.1, 1.0 and 10 $\mu$sec along the relaxation shown in a). The displacement vectors are coloured according to their magnitude with red representing displacement magnitudes greater than 1.25$\sigma$ and blue a magnitude of zero --- see linear color bar for intermediate magnitudes.}
	\label{Fig1}
\end{figure*}

At a temperature of $0.95T_{\mathrm{f}}$ the linear temperature protocol is halted and constant temperature NVT simulations are performed for 80 $\mu$sec. Fig.~\ref{Fig1}a plots the cohesive energy per atom with the thermal component removed (via a conjugate gradient relaxation), indicating an initially rapid energy drop within the first few micro-seconds followed by a slower decrease in energy involving significant fluctuations in energy. At longer timescales, relaxation appears arrested, although strong fluctuations persist. This figure also displays the corresponding icosahedral content showing an anti-correlation with the cohesive energy confirming its usefulness as a structural measure. This icosahedral content consists almost exclusively of the smaller atoms. It is emphasized that most published atomistic glass simulations employed samples relaxed at time-scales below a few hundred nano-seconds. The present work therefore gives insight into relaxation at time-scales three orders of magnitude larger than that considered in the past.

Fig.~\ref{Fig1}b is a scatter plot of the cohesive energy per atom and icosahedral content, and displays an approximately linear relation. Data is also shown for a larger sample of 32000 atoms~\cite{Derlet2017,Derlet2018}, demonstrating this linear relation is independent of the structural realization. The data-points of this larger sample are taken from relaxation occurring at a range of temperatures much lower than the $T_{\mathrm{f}}$, showing that this line represents a general relaxation trajectory of an evolving amorphous solid for this LJ-system. The increased scatter of the 80 $\mu$sec simulations originates from the smaller size of the present glass sample. 

Fig.~\ref{Fig1}c plots the atomic displacement vectors between two configurations separated by one million MD iterations (0.001 $\mu$sec) at the times 0.001, 0.01, 0.1, 1.0 and 10.0 $\mu$sec of the relaxation. These structures have all been relaxed using a conjugate gradient method to remove the effects of thermal motion, and are thus at a local energy minima. The figures show that in the early stages of the relaxation, significant atomic displacement occurs arising from multiple LSE activity within the 0.001 $\mu$sec time interval, and that as the relaxation progresses, such LSE events become increasingly rare. 

\begin{figure}
	\begin{center}
	\includegraphics[width=0.55\linewidth,trim=0cm 2cm 0cm 2cm]{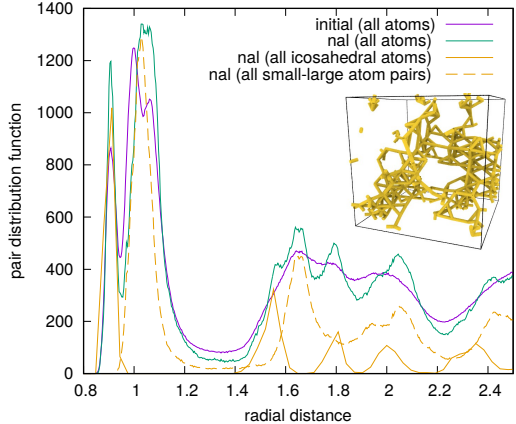}
	\end{center}
	\caption{The pair distribution function of the initial and 35 $\mu$sec structures associated with Fig.~\ref{Fig1}a. Also shown are the contributions due to only icosahedrally coordinated atoms and their neighbouring smaller atoms, of the final structure. The inset visualizes the final structure by only showing the icosahedrally coordinated atoms and the bonds between them.}
	\label{Fig2}
\end{figure}

Fig.~\ref{Fig2} plots the pair distribution function for the first configuration and the configuration at 35 $\mu$sec in Fig.~\ref{Fig1}a, revealing a weak peak-like structure beyond the nearest neighbour SRO-peak for the more relaxed structure. Such signatures of strong MRO are known to reflect icosahedral content~\cite{Sadoc1973} and are often seen in experiment~\cite{Dmowski2011,Mauro2014,Tong2015}, but rarely in simulation due to their sub-$\mu$sec relaxation time scales. The sub-PDFs indicate how this MRO arises from the spatial arrangement of the small icosahedrally coordinated atoms, and the larger atoms clustering around these. This developing spatial order is best seen by attaching nearest neighbour bonds to the icoshadrally coordinated atoms and visualizing only these bonds via OVITO~\cite{Stukowski2010}. The inset panel in Fig.~\ref{Fig2} does this and shows a structure containing repeated ring-like mosaics of Frank-Kasper polyhedra joined together by tetrahedrally packed atoms. Such structures signify the defected C15 Laves back-bone-polyhedra/crystals seen in the past works~\cite{Pedersen2010,Zemp2014,Zemp2016,Ryltsev2016} discussed in Sec.~\ref{sec-intro}. Beyond 35 $\mu$secs, the pair distribution function does not change significantly.

Thus the linearity between energy and icosahedral content seen in past work clearly extends to larger values of icosahedral content and lower values of energy --- a structural regime that includes system spanning (for our small sample) icosahedral content in the form of Frank-Kasper C15 Laves crystal fragments.

\subsection{Possible structural asymptotes} \label{ssec-ncl}

The results of sec.~\ref{ssec-lts} motivates the question as to what type of micro-structure does the relaxation embodied in Figs.~\ref{Fig1}a-b tend towards? It is again emphasized that the linear trajectory seen in Fig.~\ref{Fig1}b is also seen at temperatures significantly lower than $T_{\mathrm{f}}$~\cite{Derlet2017a}. That is, the relaxation seen at the temperature $0.95T_{\mathrm{f}}$ is the one and the same as that occurring at temperatures well below this glass transition temperature regime.

\begin{figure}
	\begin{center}
	\subfloat[]{\includegraphics[width=0.45\linewidth,trim=3cm 9.5cm 3cm 10cm]{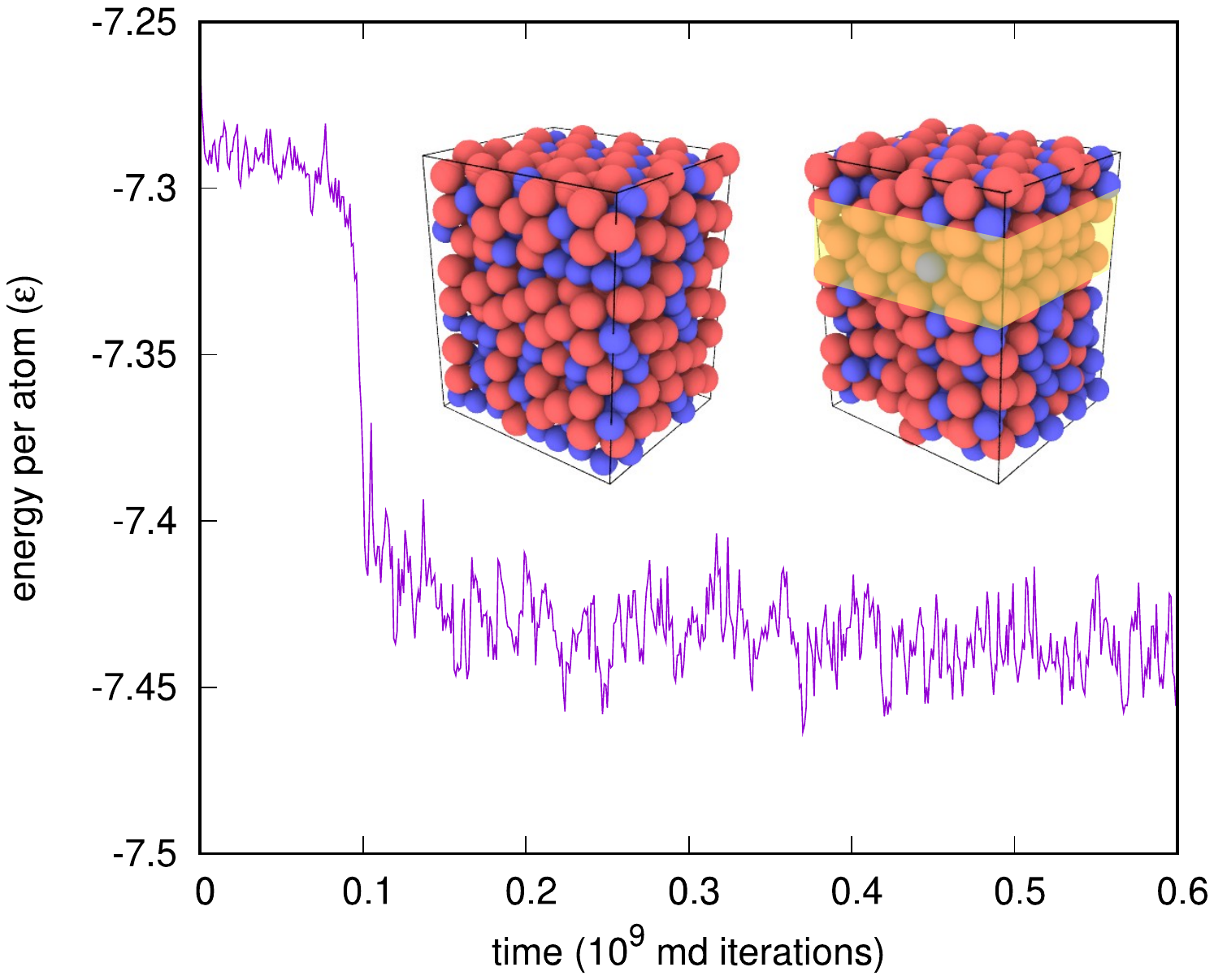}}
	\subfloat[]{\includegraphics[width=0.45\linewidth]{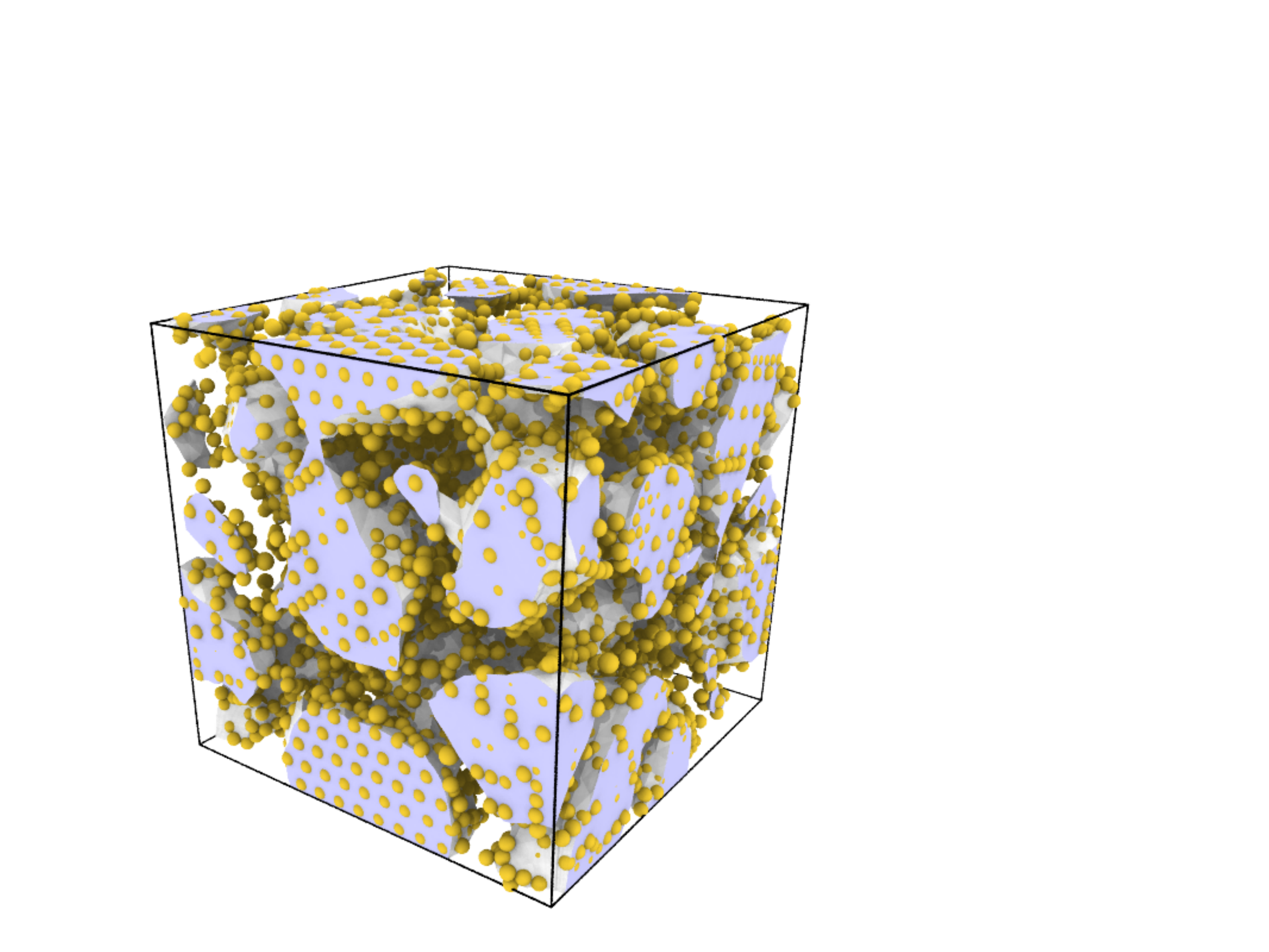}}
	\end{center}
	\caption{a) Plot of energy per atom versus hybrid MD/MC iteration number for an initially chemically disordered laves phase with a 50:50 composition of small:large atoms. The inset plots (left) the initial structure and (right) the final structure, the latter of which shows a close-packed structure consisting of only the larger atoms (high-lighted in yellow), and a less chemically disordered laves crystal with a 67:33 atom type composition. b) Geometrically constructed nano composite Laves structure with C15 crystallites embedded in a amorphous matrix existing predominantly of the larger atom. Only icosahedrally coordinated atoms are shown with a surface mesh enclosing regions containing such atoms.}
	\label{Fig3}
\end{figure}

To help answer this question, we construct a chemically disordered Laves C15 phase in which some of the larger atoms have been replaced with smaller atoms to accommodate the 50:50 composition of our present glassy system. Within the restrictions of the C15 crystallography, we choose a system size comparable to the 500 atom sample of the previous section. This ensures that any developing microstructural length-scale cannot exceed that of the well relaxed 50:50 glassy sample. To investigate the dynamics at the same temperature and volume per atom, we use hybrid MD/MC, where at every 100 MD steps we perform one MC particle exchange between atoms of different type. Doing so, eventually results in a segregation to two crystalline structures --- a Laves C15 structure with composition 67:33 and some chemical disorder, and an ordered close-packed structure consisting primarily of the larger atoms ---  see inset of fig.~\ref{Fig3}a. The dynamics of this process are mapped onto our energy versus icosahedral plot as red points in Fig.~\ref{Fig1}b. This energy and that in fig.~\ref{Fig3}a have the thermal component removed. Two clusters of points are seen, both having an icosahedral content of the order of 50\%. Such a high icosahedral percentage is expected since all small atoms are icosahedrally coordinated within the perfect Laves C15 structure. The higher energy cluster of points in fig.~\ref{Fig3}a represent the initial chemically disordered C15 structure, whereas the lower energy cluster is indicative of the final structural state the hybrid simulation reaches. 

The remarkable result in Fig.~\ref{Fig1}b is that the final cluster of data points aligns with the intersection of the relaxation trajectory of the glasses, suggesting glassy relaxation could tend towards a structure characterized by a two-phase crystalline composite consisting of Laves C15 crystals and a close-packed region of larger atoms --- with a micro-structural length-scale comparable to the current system sizes of about 500 atoms. 

Inspired by experimentally observed two-phase nano-structures~\cite{Calin2003,Pekarskaya2003,Wu2017}, a nano-structure is artificially constructed consisting of randomly orientated Laves C15 crystals embedded within an amorphous matrix of only the larger atoms. This is done by exploiting geometrical methods normally used for computer generated nanocrystalline metals~\cite{DAC2008,Froseth2005}. This procedure uses the Voronoi tessellation method to construct a Laves C15 grain network, which initially results in planar interface regions of finite thickness into which an disordered arrangement of the larger atoms are inserted at a density compatible with the global volume per atom and global 50:50 composition. These samples are then relaxed using hybrid MD/MC NVT to produce micro-structures that are stable at the $\mu$sec-timescale. The energy and icosahedral values of the final nano-composite Laves (ncL) structure are shown in Fig.~\ref{Fig1}b, for two average C15 crystallite sizes corresponding to about 400 ($r\simeq4\sigma)$ and 1200 ($r\simeq6\sigma$) atoms. It is seen that these artificially constructed candidate structures again line up with the relaxation trajectory of the corresponding glasses, for which the smaller grain sized sample has a higher energy than that of the well relaxed glass. The larger grain sized sample is however at a lower energy and higher icosahedral composition than the glass, suggesting that the associated grain size reflects a more realistic micro-structural length-scale towards which the glass relaxes. Fig.~\ref{Fig3}b visualizes this smaller grain-size sample, where only the icosahedrally coordinated atoms are shown. To aid in visualization, a surface mesh bounds the volume of these atoms, encompassing the C15 grains and a neural-like network of icosahedral connectivity between the C15 grains, which has developed during the hybrid MD/MC relaxation simulations.

Further simulations demonstrate that a simulation cell consisting only of the larger atoms at a volume per atom comparable to the average volume per atom of the amorphous interface region, rapidly crystallizes into a system-spanning closed packed structure. Thus the mono-atomic amorphous interface region of our nanophase material is stabilized by the surrounding Laves crystals, suggesting a strongly constrained glassy structure. Moreover, detailed inspection of the thermally-activated atomic displacements at $T_{\mathrm{f}}$ shows string-like excitations concentrated mainly at the interface between the Laves crystal and the amorphous region, but not within the latter.

\subsection{$T>T_{\mathrm{f}}$ relaxation and variation of chemical content} \label{ssec-lts1}

\begin{figure*}
	\begin{center}
	\subfloat[]{\includegraphics[width=0.45\linewidth]{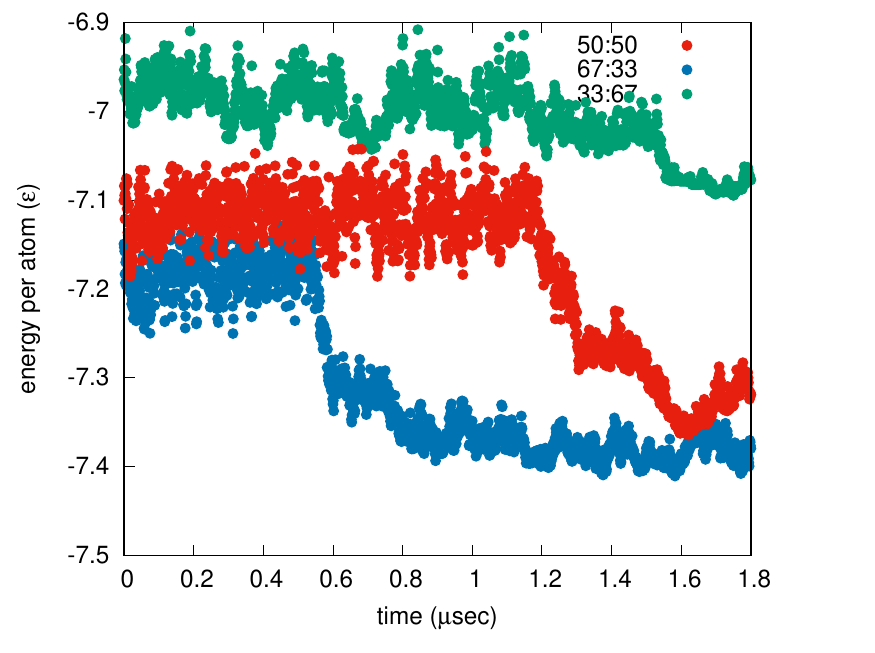}}
	\subfloat[]{\includegraphics[width=0.45\linewidth]{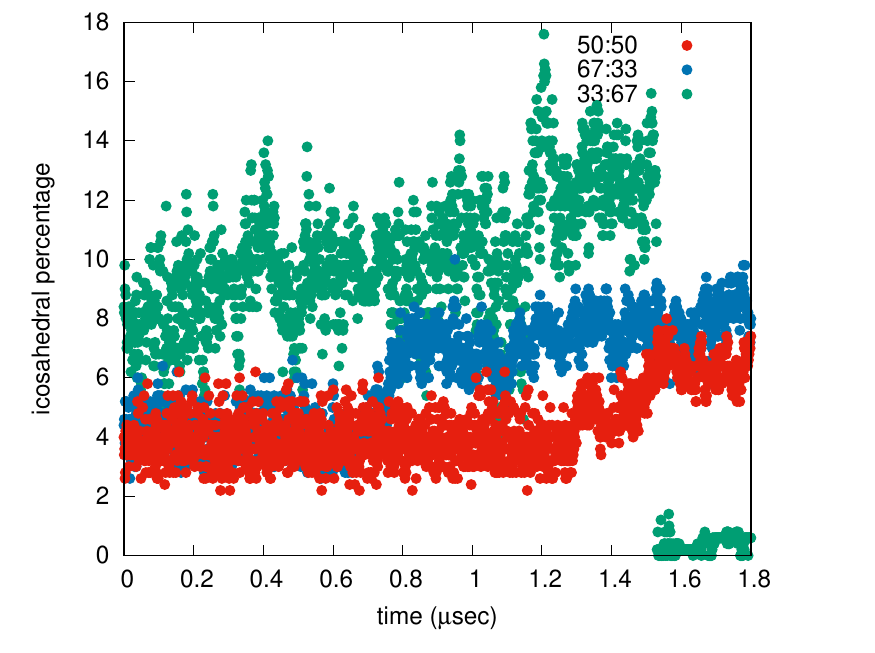}} \
	\subfloat[]{\includegraphics[width=0.45\linewidth]{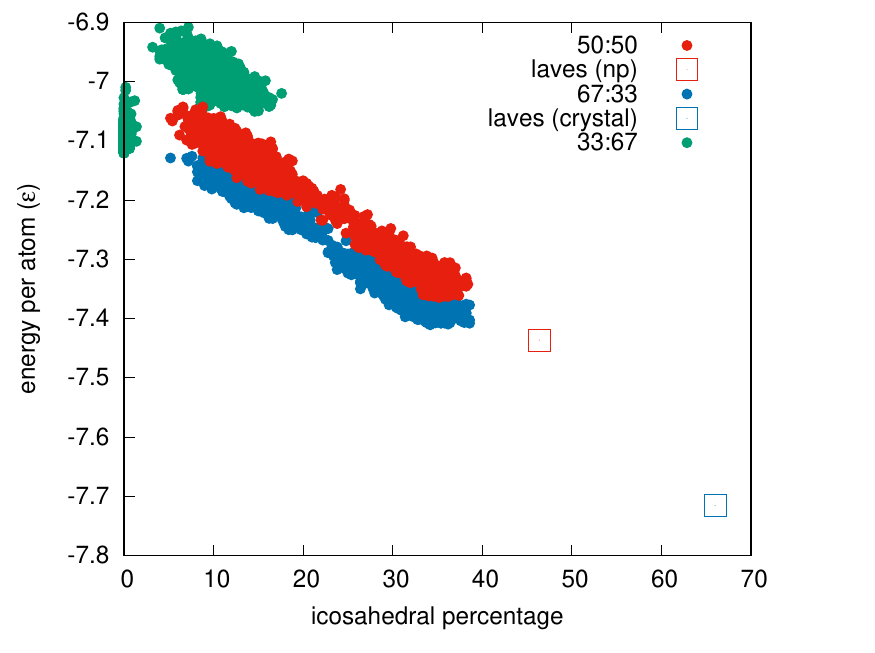}}
\end{center}
	\caption{a) Energy per atom and b) icosahedral content evolution over a time at an annealing temperature $1.2T_{\mathrm{f}}$ for the small:large atom type compositions 33:67, 50:50 and 67:33. c) Corresponding scatter plots of energy per atom and percentage icosahedral content. Also shown are the isolated data points of the perfect laves structure and the nano composite laves structure of Fig.~\ref{Fig3}b evaluated at the corresponding volumes.}
	\label{Fig4}
\end{figure*}

A similar annealing protocol for the 50:50 sample is now performed at a temperature equal to 1.2$T_{\mathrm{f}}$ (here the temperature quench from the melt has been halted at 1.2$T_{\mathrm{f}}$ instead of 0.95$T_{\mathrm{f}}$). The energy per atom and icosahedral content per atom as a function of time is plotted in Fig.~\ref{Fig4}a. At this higher temperature, initially there is little relaxation when compared to the first $\mu$sec of the lower temperature constant temperature anneal of Fig.~\ref{Fig1}a. Brief excursions to lower energy structures are however seen, and at approximately 1.2 $\mu$sec a significant irreversible reduction in structural energy occurs. The anti-correlation with icosahedral content is again seen (Fig.~\ref{Fig4}b) and the resulting relaxation trajectory is shown in Fig.~\ref{Fig4}c. Inspection of the figure reveals the same linear correlation as seen in Fig.~\ref{Fig1}b. Indeed, from a general structural perspective there is little difference between that of the final state arising from this higher temperature and that of the structure at a comparable energy per atom arising from the 80 $\mu$sec anneal, where in both cases the icosahedrally coordinated atoms arise from fragments of the C15 Laves crystal structure.

Applying the present results to a variation of chemistry, suggests that a 67:33 composition of small to large atoms would exhibit a relaxation trajectory which asymptotes to the perfect C15 Laves crystal or poly-crystal. For a direct validation of this we redo the $T>T_{\mathrm{f}}$ isotherm NVT simulations now using a 67:33 composition, but at the fixed volume for which the average volume available to the small and large atoms remains the same as that of the 50:50 simulations~\footnote{At the fixed volume of $0.76\sigma^{3}$, the average volume per atom (derived from a Voronoi tessellation) of the different sized atoms is $\Omega_{\mathrm{small}}=0.579\sigma^{3}$ and $\Omega_{\mathrm{large}}=0.955\sigma^{3}$. The global volume for an $N=N_{\mathrm{small}}+N_{\mathrm{large}}$ atom system, is then taken as  $N_{\mathrm{small}}\Omega_{\mathrm{small}}+N_{\mathrm{large}}\Omega_{\mathrm{large}}$.}. Fig.~\ref{Fig4} displays the energy and icosahedral content as a function of time, and at approximately 0.6 $\mu$sec an abrupt change occurs demonstrating significant structural relaxation. This may also bee seen in the resulting relaxation trajectory shown in fig.~\ref{Fig4}c, where also the energy and icosahedral data-point corresponding to the perfect Laves structure at the same volume per atom is shown. As with the 50:50 relaxation trajectory there is alignment between the glass and the suggested asymptotic structure, confirming our initial hypothesis of a fundamental connection between glass relaxation and the Laves crystal structure.

To further investigate the effect of stoichiometry, we also perform similar simulations for the 33:67 composition which, by our current reasoning, would result in small C15 crystallites surrounded by a dominate mono-atomic amorphous phase of the larger atom. This structural asymptote scenario is however unlikely since we have already demonstrated that the remaining mono-atomic region (without a high density of laves crystal surfaces) rapidly crystallizes to a close-packed structure. Inspection of the corresponding energy versus icosahedral trajectory in Fig.~\ref{Fig4}c does indeed demonstrate an abrupt cessation of the trajectory in terms of the icosahedral structural measure. Inspection of the atomic scale processes underlying this behaviour reveals an initial segregation of the smaller atoms into clusters involving several atoms and the creation of icosahedral content. This is followed by a system-spanning ordering to a chemically disordered cubic structure dominant in FCC/HCP content, and negligible icosahedral content (see Fig.~\ref{Fig4}b).

The similar gradients seen in Fig.~\ref{Fig4}c are a result of the appropriately scaled global volumes which ensure that in all three compositions the same average volume is available to the differently sized atoms. This reiterates the importance of local volume. When the 67:33 and 33:67 simulations are done at the fixed volume of the 50:50 simulations, similar trajectories are seen (with the 67:33 trajectory again asymptoting to the Laves crystal limit), but with gradients which now depend on composition.

\subsection{Global and local energy} \label{ss-eos}

\begin{figure*}
	\begin{center}
		\includegraphics[width=0.55\linewidth]{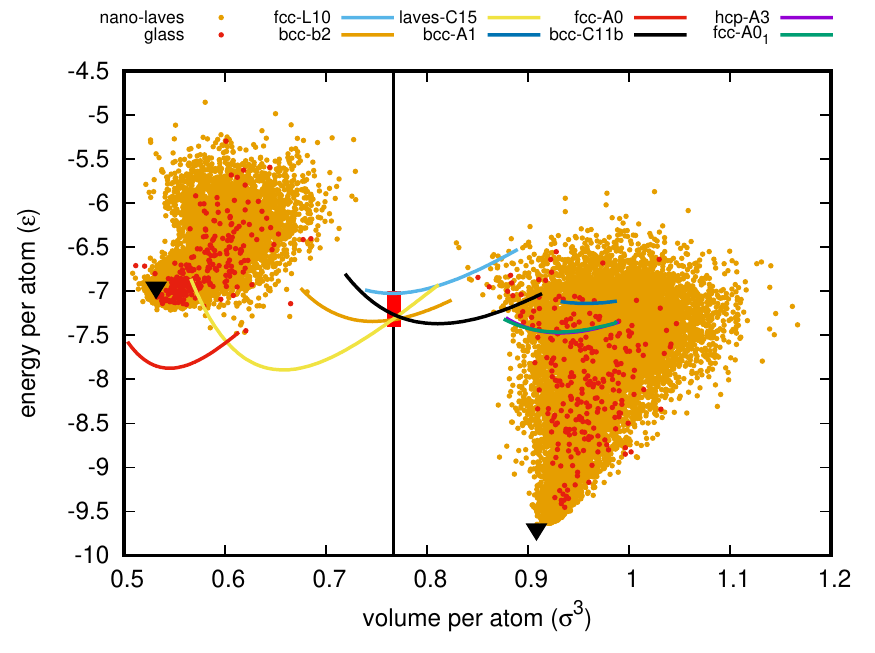}
	\end{center}
	\caption{Plot of equation-of-state curves for a number of pure and binary alloy phases. Also included are scatter plots of local potential energy versus local volume for the 80 micro-second annealed glass and the nano composite Laves structure at a fixed volume of 0.767$\sigma^{3}$ --- indicated by the black vertical line, the thicker red part of which indicates the energy per atom range of the glass. The two inverted black triangles indicate the volume and energy per atom of the equilibrium laves C15 phase at a similar fixed volume.}
	\label{Fig5}
\end{figure*}

To gain a better understanding of the energetic origin of the observed behaviour, Fig.~\ref{Fig5} displays the equation of state (EOS) curves for the Wahnstr{\o}m LJ potential for a variety of relevant crystal phases. The black vertical line indicates the fixed volume at which the present simulations were performed, and it is seen that the Laves, the intermetallic bcc-b2 and the bcc-C11b phases have comparable energies at this chosen volume per atom. Note that there is also little difference between the EOS curves of the hcp and fcc phases of each atom type, which is to be expected for a short-range pair potential.

The Laves C15 EOS curve indicates that a significant reduction in energy could be achieved by forming this phase at a reduced volume, and that the remaining larger atoms of non-stoichiometric compositions (such as 50:50) could become hcp/fcc by increasing their volume per atom. This possibility may be quantified by writing the energy of this dual phase as
\begin{equation}
E_{\mathrm{Laves-hcp}}=\frac{3}{4}E_{\mathrm{Laves}}\left(V_{\mathrm{Laves}}\right)+\frac{1}{4}E_{\mathrm{hcp}}\left(V_{\mathrm{hcp}}\right) \label{Eqn1}
\end{equation}
where the volume per atom of each phase must satisfy the global fixed volume per atom, V,
\begin{equation}
V=\frac{3}{4}V_{\mathrm{Laves}}+\frac{1}{4}V_{\mathrm{hcp}}. \label{Eqn2}
\end{equation}
Using the EOS curves of Fig.~\ref{Fig5}, a minimum of Eqn.~\ref{Eqn1} is found to be $-7.7\varepsilon$ which occurs at $V_{\mathrm{Laves}}=0.69\sigma^{3}$ and $V_{\mathrm{hcp}}=1.00\sigma^{3}$. Such an energy does not include the cost of the interface energy and therefore represents the limit where the microstructural length-scales have diverged and the interface region becomes negligible compared to the volume occupied by the Laves and hcp/fcc phases. Indeed, drawing a line between the energies of the two ncL micro-structures (which are distinguished by the two grain sizes) and extrapolating this to the regime of an icosahedral content of $50\%$ yields an energy close too and slightly higher than the bulk value of $-7.7\varepsilon)$. Together, these results give insight into the origin of the energetics of the dynamics seen in Fig.~\ref{Fig1}b. Whilst certainly influenced by the bulk energies, the interface energy between the laves or laves-like regions and an approximately mono-atomic close-packed or amorphous structure plays a significant role in setting the overall micro-structural energy scale and therefore the emergent length scales of the well relaxed amorphous solid. 

Further insight can be gained by inspecting the distribution of local energies and volumes in the considered systems. For the perfect Laves phase, the local energy and volume are indicated by the black inverted triangles in Fig.~\ref{Fig5} with the low/high volumes representing the small/large atoms. Also included in Fig.~\ref{Fig5} is a scatter plot of the local energies of the final glass structure of the 80 $\mu$sec anneal, and of the artificially constructed ncL microstructure (grain size $6\sigma$). Both structures show local energies that are centered close to the equilibrium local energies of the perfect Laves crystal --- this becomes more clear when a density plot with respect to local energy and volume is made (not shown). This is particularly the case for the smaller atoms which are predominantly in icosahedral environments. The remaining atoms not associated with the Laves structures tend to have higher energies and volumes, especially the larger atoms in the amorphous interface of the ncL microstructure. Thus the local energies and volumes of the amorphous structure, derived from a direct quench from the melt and a long-time anneal are similar to the artificially generated nano-phase Laves structure generated in Fig.~\ref{Fig3}b.

\section{Discussion} \label{sec-discussion}

\begin{figure*}[t]
	\begin{center}
		\includegraphics[width=0.9\linewidth,trim=0cm 4cm 2cm 4cm]{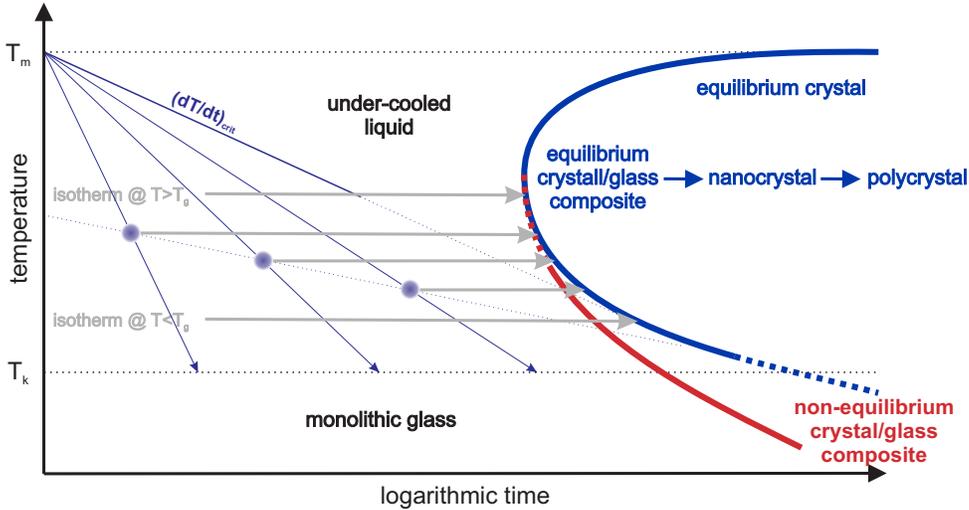}
	\end{center}
	\caption{Schematic of a temperature-time-transformation (TTT) diagram showing the glass and crystalline regions which can be accessed during a temperature quench protocol from the liquid state. The region indicated by red represents a possible intermediate regime in which out-of-equilibrium nano composite structures are encountered that are not consisting of the equilibrium polycrystalline phases of the inner nose region.}
	\label{Fig6}
\end{figure*}

The temperature-time-transformation (TTT) diagram provides a framework in which to quantify when either glass formation or crystallization occurs. Fig.~\ref{Fig6} details a schematic of such a diagram with a normalized temperature regime on the vertical axis. The quantitative details depend strongly on the chemistry of the melt. Traditional quenching of a glass-forming metallic liquid requires cooling rates less than the critical value $(dT/dt)_{\mathrm{crit}}$, such that the temperature-time trajectory does not intersect the so-called nose of the crystalline region. Exceptional glass formers have low $(dT/dt)_{\mathrm{crit}}$ of the order of 1 K/s, whereas mono-atomic metallic glasses may require quench rates up to 10$^{14}$ K/s to bypass the crystallization nose~\cite{Mao2014}.

The crystallization nose in Fig.~\ref{Fig6} extends over a finite time interval, underlying a transition (as time evolves) from a dominant glassy structure to a dominant equilibrium polycrystalline structure. On the glassy side of the nose this encompasses the micro-structural regime of nanocrystalline composites ~\cite{Calin2003,Pekarskaya2003,Wu2017} --- a microstructure characterized by nanocrystals of the equilibrium crystalline structure typically embedded in a glassy matrix. To follow a particular isotherm above the glass transition temperature regime, after a heating or quenching, is experimentally challenging due to the short accessible time-window prior to the crystallization nose, and requires exceptionally stable glassy alloys with a crystallization nose located at several tens of seconds~\cite{Loffler2000}. Alternatively, new high-speed calorimetric instrumentation can now be used to access the shorter time-scales of a TTT diagram~\cite{Schawe2019}. 

The consequences of following such isotherms are best illustrated for the two different regimes, $T<T_{\mathrm{f}}$ and $T>T_{\mathrm{f}}$. The first case considers the structural trajectory of a glassy solid, as done in the simulations of sec.~\ref{ssec-lts}. The typical timescales associated with MD places the onset of the $\mu$sec MD isotherm far to left of the TTT diagram. At the beginning of the isotherm, the glass is unrelaxed and the connectivity of SRO is low. With increasing time, the icosahedral content increases, leading to the development of MRO and internal length scales which increase as a function of annealing time. The second isotherm in Fig.~\ref{Fig6} at $T>T_{\mathrm{f}}$, is motivated by the idea that shorter time-scales are needed to approach the crystallization nose. The results of sec.~\ref{ssec-lts1} indeed demonstrate this, where within the order of 1$\mu$sec, significant relaxation associated with the creation of icosahedral SRO and MRO is seen, producing structures not so different from the 80 $\mu$sec anneal of sec.~\ref{ssec-lts} at $T<T_{\mathrm{f}}$ via a similar relaxation trajectory (Figs.~\ref{Fig1}b and \ref{Fig4}c).

The existence of glassy microstructural length scales, typically referred to as spatial heterogeneities, has been demonstrated in numerous experiments at both the nanometer and micrometer scale~\cite{Liu2018,Qiao2019,Ross2017,Liu2011}, but their structural origin remains unclear. Based on Figs.~\ref{Fig1}, \ref{Fig4}c and~\ref{Fig6} it becomes apparent that numerous final glassy or nc materials can be realized by isothermal processing and varying initial composition. Spatial heterogeniety length-scales can be tuned along an isotherm, and are expected to maximize at a length-scale set by the forming nc-phase. Whilst only shown for binary model glass formers, it could be that recent reports of unusual nano-phase metallic glass matrix composites~\cite{Rupert2015,Wu2017} are a result of the here traced isothermal evolution of an out-of-equilibrium two-phase structure, where no immediate crystalline phase is available to all constituent elements of the glass. Thus, the relaxation into a nc phase is expected to be less probable with increasing chemical complexity, since this increases the probability for the remaining amorphous matrix --- now at a different composition than the initially glass forming system --- to find crystallization pathways. The mechanisms discussed here should be distinguished from the micro-structural length-scales arising in eutectic-like compositions for which segregation effects can occur due to spinoidal decomposition in the under-cooled liquid state~\cite{Zhao2014}.

For the geometrically constructed nc Laves micro-structure (Fig.~\ref{Fig3}b), the surrounding grain structure appears to stabilize the amorphous matrix. In this context it is noteworthy that out-of-equilibrium processing of binary CuZr alloys via ball milling and subsequent annealing has produced nc structures~\cite{Rupert2015} that may form due to the reasons discussed here. In this case, an isothermal annealing at 90$\%$ of the melting temperature lead to the formation of a stable glassy interface. The origin of this experimental observation can now be framed within a chemical segregation effect that promoted the formation of inter-grain compositions not able to form a thermodynamically favored crystalline phase. In fact, the original work reporting nanometre-scale glassy boundaries in nanocrystalline CuZr argued that an enrichment of Zr-atoms in the boundary regions promoted its formation~\cite{Rupert2015}. This is very much in agreement with the here observed amorphous matrix that forms via the complete depletion of the smaller atoms for the equiatomic composition.

The observation that the relaxation trajectory seen in Fig.~\ref{Fig1}b occurs at temperatures well below $T_{\mathrm{f}}$ (See refs.~\cite{Derlet2017a,Derlet2018}) as well as temperatures above $T_{\mathrm{f}}$ (fig.~\ref{Fig4}), and that the trajectory is inclusive of geometrically constructed Laves nano-composite structures, suggests that in the low temperature amorphous solid regime, structural relaxation is also driven by the formation of an extended network of Laves back-bone structures which might eventually lead to Laves crystallite structures. This would lead us to a well-relaxed glassy state which is not fully amorphous and has structural length-scales well above that of the SRO and traditional MRO scale.

The appropriate historical context lies in the seminal work of Kauzmann~\cite{Kauzmann1948} which was concerned with the low temperature limit of the under-cooled liquid and predates the potential energy landscape picture of Goldstein~\cite{Goldstein1969}. As the temperature reduces the non-vitrified meta-stable liquid increasingly exhibits structural configurations that are neither representative of the equilibrium crystal nor of the equilibrium liquid --- the so-called ``ambiguous regions'' of phase space. A key concept was that these ambiguous regions entail dynamics in which the timescales associated with equilibrium crystal nucleation and atomic mobility converge. The finite temperature at which this occurs characterizes the Kauzmann temperature, $T_{k}$, at which the configurational entropy would approach that of the crystal and become sub-extensive. Three structural scenarios were envisaged: 1) the equilibrium crystal structure emerges thus entirely avoiding the first order nature of the equilibrium liquid-to-solid transition, 2) the emergence of a new state of ``high order'' and 3) an amorphous structure with a sub extensive configurational entropy. Scenario 1) was dismissed since experiment suggested that the free energy of the meta-stable under-cooled liquid does not converge to that of the relevant equilibrium crystal phase (or phases). Scenario, 2), was not considered viable since it was difficult to conceive of a ``plausible structure''. Kauzmann ultimately chose 3) entailing that a glass was indistinguishable from the liquid at $T_{k}$ --- the ideal glass.

If one assumes that structural relaxation below $T_{\mathrm{f}}$ leads to Laves crystallites, the present work suggests two possible variants of scenario 2): i) the emerging Laves C15 phase is an equilibrium crystal phase of the empirical potential for the atom-type composition considered, or ii) it is a non-equilibrium crystal phase. 

If it is ii), then the TTT diagram will need to distinguish between nc regimes consisting of equilibrium or out-of-equilibrium crystallite phases. Indeed, one could envisage that the equilibrium regime would dominate at higher temperatures, $T>T_{\mathrm{f}}$ and towards the melting temperature regime, whereas the latter becomes relevant at lower temperatures, $T<T_{\mathrm{f}}$ and close to the glass transition temperature regime. The scenario of the Laves phase being non-equilibrium is interesting since it suggests the ``high order'' structure of Kauzmann has little to do with the relevant equilibrium crystal phase(s), but rather is intimately connected to the reduction of frustration associated with glass relaxation as suggested by the works of Frank and Kasper~\cite{Frank1952,Frank1958}, Chaudri and Turnbull~\cite{Chaudhari1978}, and Nelson and co-workers~\cite{Nelson1983a,Nelson1983b}. If, however, it is i) and the Laves C15 phase is an equilibrium crystal phase, then the structure of ``high order'' nano-composite partially involves an equilibrium crystal phase (mixed with a glassy interface region) and no such modification of the TTT diagram would be needed.

Whilst the binary phase diagram of the LJ Wahstr{\o}m potential is not known, the zero-temperature equation of state curves displayed in Fig.~\ref{Fig5} and discussed in sec.~\ref{ss-eos} indicate a purely energetic reason for the formation of the Laves crystal phase. Free-energy equation of state curves will ultimately be needed to answer whether or not this is the equilibrium phase, however it is noteworthy to point out that a similar energetic reason for the creation of a Laves phase is seen for the embedded atom method CuZr potential when producing analogous equation of state curves (not shown) ---  a material specific model system that also exhibits the growth of Laves fragments~\cite{Zemp2014,Zemp2016,Ryltsev2016}. For these simulations, there exists an experimental counterpart for which the binary phase diagram is precisely known~\cite{Okamot02009}. For compositions close to the inter-metallic 50:50 composition, the equilibrium phase has little to do with the Laves phase and much more to do with the B2 BCC intermetallic phase. On the other hand, for Eutectic-like compositions near 67:33, the Laves C15 structure can be an equilibrium phase. Thus whether or not the TTT diagram needs to be modified to include a nano-composite regime involving non-equilibrium structures will depend on the chemical content. Of course the used CuZr empirical potential might be unable to describe this regime of equilibrium behaviour correctly, but it is encouraging that there exists experimental evidence for the formation of Laves fragments in the CuZr binary glass system~\cite{Wang2008,Li2009,Hwang2012}. 

Alternatively, one could envisage that structural relaxation below $T_{\mathrm{f}}$ does not lead to Laves C15 crystallites within the experimental timescale, but rather to a disordered and interconnected collection of Frank-Kasper backbone structures with a strong signature of MRO --- the possible asymptote to the ideal glass. In other words kinetic arrest below $T_{\mathrm{f}}$ limits the accessible regime of the relaxation trajectory of Fig.~\ref{Fig1}b and \ref{Fig4}c. In this scenario, the TTT diagram could again be modified to represent the emerging structural length scales associated with this asymptote, indeed the nano-composite microstructural picture might still have relevance, although now based on the details of an entirely non-crystalline structure. Indeed the possibility of such kinetic arrest provides a more precise formulation of questions concerned with the constraints inherent to a glass structure --- a theme discussed in Sec.~\ref{sec-intro}. In particular, do such constraints (what ever they are) preclude glassy structural evolution that eventually ends with the formation of Laves crystallites? Using the language suggested by Nelson and co-workers, is the disorded disclination network of a well-relaxed glass able to transform to the ordered disclination network of a topological close-packed structure, such as the C15 phase? Might the amorphous solid be topologically protected from such evolution? 

\begin{figure*}[t]
	\begin{center}
		\includegraphics[width=0.9\linewidth,trim=2cm 2cm 2cm 2cm]{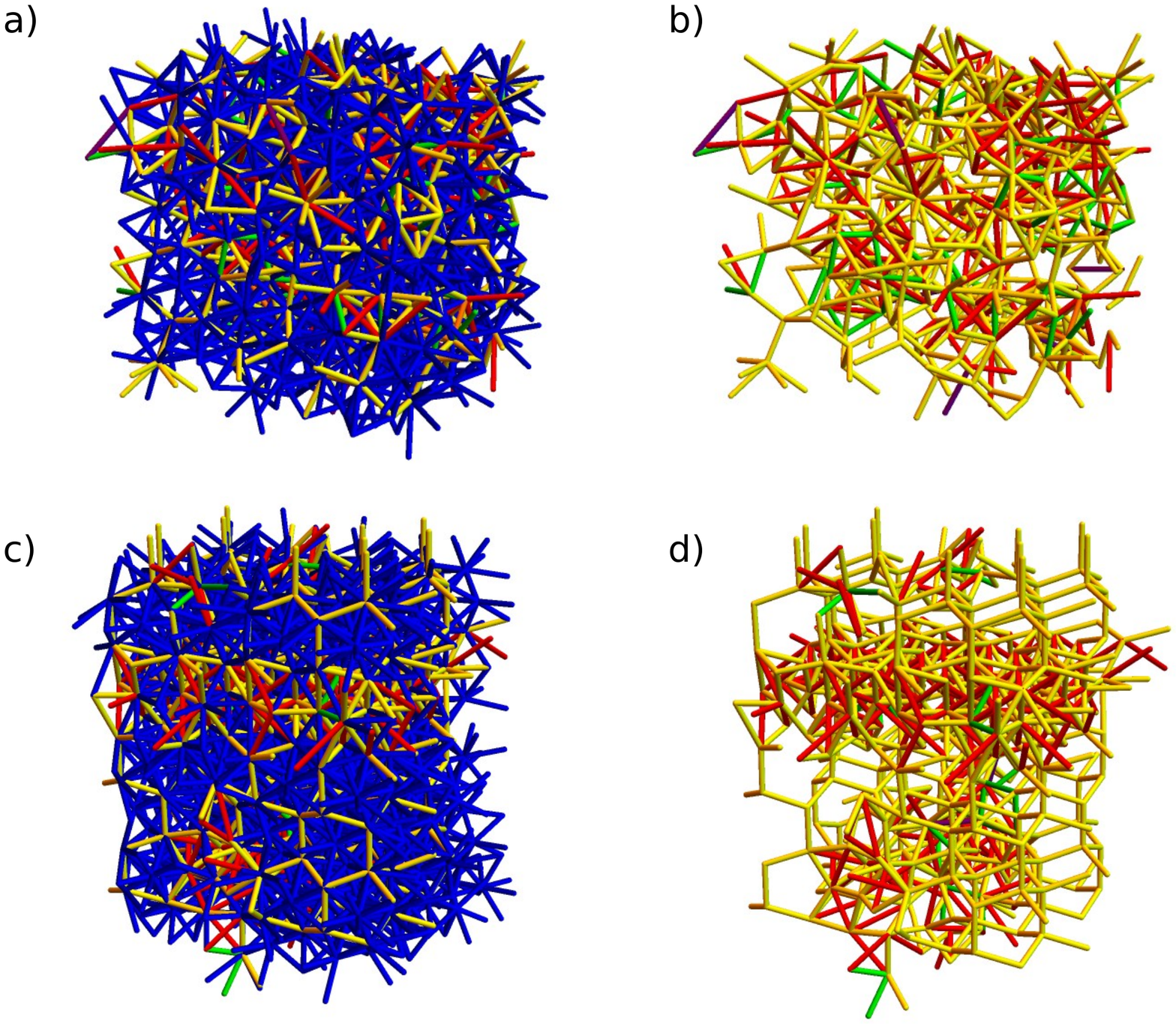}
	\end{center}
	\caption{Visualization of nearest neighbour bond network for the well relaxed glass (top panels) and the well relaxed laves/closepacked crystal structure of Fig.~\ref{Fig3}a (bottom panels). Here each bond is coloured according to the local defect disclination structure, where purple/red/blue/yellow/green/orange represent $n$-fold bonds with $n=3/4/5/6/7/8$. For both structures, the right hand panels (b and d) omits the, disclination free, 5-fold symmetric bonds. The three layers in c) containing a low density of blue bonds, represents the close-packed structure of the larger atoms.}
	\label{Fig7}
\end{figure*}

Figs.~\ref{Fig7}a plots the nearest neighbour bond network of the final configuration of the relaxation entailed in Fig.~\ref{Fig1}a. Each bond is colored according to the number of common neighbours, $n$, of the two atoms that make up the bond, and may be referred to as either a $n$-fold symmetric bond --- corresponding to the number of distorted tetrahedrons packed around the bond. In the figure the blue bonds represent the five symmetric/disclination free bonds, whereas the yellow bonds represent the six-fold/disclinated bonds underlying the close-packed structure. Fig.~\ref{Fig7}b omits the disclination free bonds and would in principle represent the disclination network of the glass structure --- this should be compared to figure 3 of Ref.~\cite{Nelson1983b} which is a schematic of a glass containing significantly less disclination structures. The figures show a disordered network of blue and yellow bonds reflecting the glass's MRO and network of Frank-Kasper backbone polyhedra. Fig.~\ref{Fig7}c-d similarly visualize the approximate Laves/close packed structure displayed in Fig.~\ref{Fig3}a. Inspection of these figures reveals an extended and ordered network of both five-fold symmetric and six-fold symmetric bonds within the Laves crystal regime. 

The work of Nelson~\cite{Nelson1983b,Mermin1979,NelsonBook} indicates that the algebra corresponding to the connectivity of the glassy disclination network is non-abelian, which entails strong constraints on the available disclination configurations to the extent that the entanglement seen in Figs.~\ref{Fig7}a-b (say) might be topologically protected against structural evolution towards long-range order. This should be contrasted with the dislocation network of a crystal, whose defect algebra is abelian (additive) and thus far less constrained in terms of the available configurations structural evolution could lead to. This perspective is certainly an interesting line of further research.

\section{Summary and concluding remarks} \label{sec-conc}

Molecular dynamics have been used to investigate model binary glass relaxation at temperatures above and below the glass transition temperature regime. The approximately linear relation between energy per atom and the fraction of atoms in a local icosahedral environment is found to extend to higher icosahedral compositions and therefore to significantly more relaxed amorphous structures. We identify this relation as a relaxation trajectory that:
\begin{enumerate}
\item occurs during isothermal protocols both above and below the glass transition temperature regime.
\item reflects a structural evolution in which the medium range order increases through a growing population of inter-connected Frank-Kasper polyhedra fragments.
\item apart from an energy shift, remains the same for a range of chemical compositions when the atom types have similar accessible characteristic volumes.
\item encompasses geometrically constructed nano-composite microstructures consisting of Laves C15 crystallites of a particular size embedded in an amorphous matrix consisting primarily of the larger atoms, and a micro-structural energy and length scale that is determined also by the interface energy.
\end{enumerate}
The mono-atomic amorphous interface region of the nano-composite is found to be strongly constrained, and therefore stabilized, by the surrounding Laves C15 crystallites. 

Taken together, the above trajectory implies a structural relaxation path of a binary glass that naturally leads to emergent structural length-scales, or spatial heterogeneities which might involve the crystallization of the Laves phase, or more generally topological close-packed phases. Such crystal structures are dominated by five-fold symmetric bonds which are closely associated with the local minimization of structural frustration within the glass. Thus the evolution of such spatial heterogeneities may now be understood as a manifestation of medium-range-ordered evolution that might eventually lead to nanometer-sized crystalline structures. Alternatively, via fig.~\ref{Fig1}b, the emerging spatial heterogeneities will strongly correlate with the amount of energy stored within the glassy structure. This suggests, a regime of so-far unidentified and unexploited nano-composite structures, hidden in the lower-temperature regime below the nose in a temperature-time-transformation diagram, which will be a sensitive function of binary (or ternary) composition, annealing protocol and the stability of any remaining glassy matrix phase.

\section{Acknowledgments}

The present work was supported by the Swiss National Science Foundation under Grant No. 200021-165527

\end{document}